\documentclass[a4paper,11pt,final]{article}

\usepackage{authblk}

\usepackage{graphics}
\usepackage{graphicx}
\usepackage{epsfig}

\usepackage{amssymb}
\usepackage{amsthm}

\date{}

\begin{document}
%\begin{frontmatter}



\title{One-dimensional hyperbolic transport: positivity
and admissible
boundary conditions derived from the wave formulation}


\author[1]{Antonio Brasiello}
\author[2]{Silvestro Crescitelli}
\author[3]{Massimiliano Giona$^*$}
\affil[1]{Dipartimento di Ingegneria Industriale
Universit\`{a} degli Studi di Salerno
via Giovanni Paolo II 132, 84084 Fisciano (SA), Italy }
\affil[2]{Dipartimento di Ingegneria Chimica, dei Materiali e della Produzione Industriale 
Universit\`{a} degli Studi di Napoli ``Federico II'' 
piazzale Tecchio 80, 80125 Napoli, Italy}

\affil[3]{Dipartimento di Ingegneria Chimica DICMA 
Facolt\`{a} di Ingegneria, La Sapienza Universit\`{a} di Roma 
via Eudossiana 18, 00184, Roma, Italy 
\authorcr $^*$ massimiliano.giona@uniroma1.it}

\maketitle

\begin{abstract}
We consider the one-dimensional Cattaneo equation for transport of
scalar fields such as solute concentration and temperature in
mass and heat transport problems, respectively. 
Although the Cattaneo equation admits a sto\-chastic interpretation -
at least in the one-dimensional case - 
negative concentration values  can occur in boundary-value problems on a finite
interval. This phenomenon stems from the  probabilistic
nature of this model: the stochastic
interpretation provides constraints on the admissible boundary conditions,
as can  be deduced from the wave formulation here presented.
Moreover, as here shown, energetic inequalities and the dissipative nature of the equation
provide an alternative way to derive the same constraints on
the boundary conditions derived by enforcing positivity.
The analysis  reported is also extended to transport problems in the presence
of a biasing velocity field.
Several general conclusions are drawn from this analysis
that could be extended to the higher-dimensional case.

\end{abstract}

%\begin{keyword}

%\end{frontmatter}

\section{Introduction}
\label{sec_1}

Cattaneo transport equation for scalar fields (matter, heat) stands as
a valuable attempt to overcome the intrinsic problems of
Fickian transport, namely its infinite velocity of propagation \cite{cattaneo}
(throughout this article we refer  as ``Fickian'' to
the wealth of transport problems where the flux of the transported entity
is proportional to the concentration gradient, such as in the
Fick's equation for mass transport, or to temperature gradient, such as
 in the Fourier equation
for conductive heat transfer).

This is achieved by introducing an exponentially decaying memory contribution
in the flux-concentration constitutive equation.
The Cattaneo model finds important applications in some
specific transport problems (such as the phenomenon of the
second sound in superfluid helium)
\cite{sec_sound} (see also the review by Joseph and Preziosi \cite{preziosi,preziosi2}),
and in a variety of other, more engineering-oriented, studies \cite{eng1,eng2,eng3}.
Some authors have  developed an approximate mesoscopic derivation
of Cattaneo-like transport equations in the form of a generalized
hyperbolic Fokker-Planck equation \cite{mesoscopic,hyperbolic}, and Cattaneo
model rests as a fundamental example of generalized constitutive
equation in the theory of extended irreversible thermodynamics
\cite{eit1,eit2,eit3}.

Its implications in theoretical physics should not be underestimated,
as it represents a valuable contribution for extending
transport models in a relativistic (Lorentz-invariant) way \cite{hanggi},
and it constitutes the starting point for attempting a stochastic
interpretation of the Dirac equation \cite{kac1,dirac1}.

 However, its validity has been
deeply questioned  and criticized for space dimension greater than one, as the
associated Green function attains negative values \cite{criticism}, so that
the propagation of generic non-negative initial conditions can
return negative concentration values.

In  point of fact, even this observation can be pushed forward,
since there exist one-dimensional problems on the finite interval
for which the solution of the Cattaneo equation 
in the presence of non-negative initial and boundary conditions
can become negative.
This is shown in Section \ref{sec2}
by means of a simple, closed-form example.
The Cattaneo equation on the interval has been studied also
in \cite{weiss1,weiss2}, for boundary conditions
that do not present problems as it regards positivity.

This result may seem in contradiction with the stochastic
derivation of the Cattaneo model starting from
a simple stochastic equation driven by dichotomous Poisson noise,
developed earlier by S. Goldstein \cite{goldstein}
and subsequently  re-elaborated by M. Kac in a simple but seminal contribution
\cite{kac}. This contradiction is however purely
apparent and can be resolved by formulating the transport problem
in the wave-formalism envisaged by Kac \cite{kac} and subsequently elaborated
by several others authors \cite{weiss2,kolesnik1,kolesnik2,haba}, just to cite some
relevant
contributions in the field.
The wave analysis  indicates  unambiguously that not all of  the
 boundary conditions, that are commonly applied in a Fickian framework,
are
physically admissible in  hyperbolic transport schemes.
More precisely, the wave-like nature of the  transport models with memory
dictates bounds and constraints on the admissible boundary conditions
preserving positivity. This is shown in Sections \ref{sec3} and \ref{sec4}
dealing with the desorption experiment addressed in Section \ref{sec2}.
The same constraints derived from positivity emerge from the analysis
of the L$^2$-norms, i.e., from energetic conditions, once
dissipation is enforced (see Section \ref{sec5}).

In this paper the analysis is also extended to one-dimensional problems
in the presence of a deterministic biasing  velocity field,
deriving constraints on boundary conditions in the classical
problem of boundary-layer polarization characterizing transport
across permeable or perm-selective membranes (see Section \ref{sec6}).
Some general conclusions, oriented
 towards a rigorous formulation
of hyperbolic transport models,  are addressed in the concluding Section \ref{sec7}.

\section{The Cattaneo equation on the unit interval}
\label{sec2}

Consider a transport problem for a  conserved scalar field $u(x,t)$
on the spatial interval $x \in [0,L]$ and time $t$, in the absence of source terms.
In this case, the balance equation is obviously given by
\begin{equation}
\frac{\partial u(x,t)}{\partial t}= - \frac{\partial J(x,t)}{\partial x}
\,,
\label{eq2_1}
\end{equation}
and assume that the flux $J(x,t)$ is related to $u(x,t)$ via a constitutive
equation of Cattaneo type
\begin{equation}
J(x,t)= - k \frac{\partial u(x,t)}{\partial x}-\tau_c \frac{\partial 
J(x,t)}{\partial
 t} \, ,
\label{eq2_2}
\end{equation}
where $k$ is the ``Fickian'' diffusivity (conductivity) and
$\tau_c$ the characteristic memory time. By substituting eq. (\ref{eq2_2}) into eq. (\ref{eq2_1}) 
we obtain 
\begin{equation}
\frac{\partial u(x,t)}{\partial t} + \tau_{c} \frac{\partial^2 u(x,t)}{\partial t^2}=
k \frac{\partial^2 u(x,t)}{\partial x^2} \; , 
\label{eq2_cattaneo}
\end{equation}
namely the Cattaneo equation.

As $\tau_c \rightarrow 0$, the  Cattaneo equation degenerates into a
diffusion equation, while as $\tau_c \rightarrow \infty$, keeping $k/\tau_c=\mbox{constant}$, it reduces to a one-dimensional wave equation.
Both these two limit conditions, i.e., the parabolic limit and the
pure dissipation-free wave equation, bring physical contradiction
inside, namely the infinite velocity of propagation characterizing Fickian transport on one hand, and the occurrence of pure wave-like propagation
without dissipation on the other hand,
as mentioned by Joseph and Preziosi \cite{preziosi,preziosi2}.

As initial condition at $t=0$ let
\begin{equation}
u(x,0)=u_{\rm in}=\mbox{constant} \, , \qquad \left .
\frac{\partial u(x,t)}{\partial t} \right |_{t=0}=0 \; .
\label{eq2_3}
\end{equation}
Introducing the nondimensional variables $y=x/L$, $\theta=t k/L^2$,
$\alpha=\tau_c k/L^2$,  $\phi=u/u_{\rm in}$, the balance
equation for $\phi(y,\theta)$ reads
\begin{equation}
\frac{\partial \phi}{\partial \theta} + \alpha \frac{\partial^2 \phi}{\partial \theta^2}= \frac{\partial^2 \phi}{\partial y^2} \, ,
\label{eq2_4}
\end{equation}
and $\phi|_\theta=1$, $\partial \phi/\partial \theta |_{\theta=0}=0$.

Let us consider a desorption experiment, meaning that $u(x,t)$
can be viewed, for instance, as the concentration of a solute inside some
solid matrix where it diffuses. Suppose that the boundary at $x=0$,
($y=0$),  is perfectly impermeable to transport so that $J|_{x=0}=0$
for all $t>0$. Conversely, at $x=L$, ($y=1$), solute diffuses out of
the solid matrix onto an external reservoir that is much larger than
the sample matrix and perfectly mixed so that the concentration
at $y=1$ can be assumed vanishingly small, i.e.,
\begin{equation}
\phi(1,t)=0 \; .
\label{eq2_5}
\end{equation}
It follows from the constitutive equation, that the zero-flux
boundary condition at $y=0$ reduces to
\begin{equation}
\left . \frac{\partial \phi(y,\theta)}{\partial y} \right |_{y=0}=0 \; .
\label{eq2_6}
\end{equation}
From eqs. (\ref{eq2_5})-(\ref{eq2_6}) the concentration
field $\phi(y,\theta)$ can be expanded in Fourier series
\begin{equation}
\phi(y,\theta)= \sum_{n=1}^\infty \phi_n(\theta) \, \cos(\nu_n y) \; ,
\qquad \nu_n=\frac{(2 n -1) \pi}{2} \; .
\label{eq2_7}
\end{equation}
The Fourier coefficients $\phi_n(\theta)$ satisfy the equation
\begin{equation}
\alpha \, \ddot{\phi}_n(\theta) + \dot{\phi}_n(\theta) + \nu_n^2 \phi_n(\theta) =0 \, ,
\label{eq2_8}
\end{equation}
where $\dot{\phi}_n$ indicates the derivative with respect to $\theta$
and
$\ddot{\phi}_n$ the second derivative,
equipped with the initial condition $\phi_n(0)= 2 (-1)^{n-1}/\nu_n= \phi_{n,0}$,
$n=1,2,\dots$ .
The characteristic equation associated with eq. (\ref{eq2_8})
is readily $\lambda^2 + \lambda/\alpha + \nu_n^2/\alpha=0$, and
it admits the solutions
\begin{equation}
\lambda = - \frac{1}{2 \alpha} \pm \frac{\sqrt{\Delta_n}}{2 \alpha} \; ,
\qquad \Delta_n= 1- 4 \nu_n^2 \alpha \; .
\label{eq2_9}
\end{equation}
Depending on the sign of the discriminants $\Delta_n$ two cases occur:
(i) for $n \leq n^*$, where $\Delta_{n^*}>0$ and $\Delta_{n^*+1}<0$, the
roots of the characteristic equation are real: $\lambda_{1,n}=-1/2 \alpha
+ \sqrt{\Delta_n}/2 \alpha$, $\lambda_{2,n}=-1/2 \alpha
- \sqrt{\Delta_n}/2 \alpha$. Therefore, for $n \leq n^*$,
\begin{equation}
\phi_n(\theta) = A_n e^{\lambda_{1,n} \theta} + B_n  e^{\lambda_{2,n} \theta} 
\; .
\label{eq2_10}
\end{equation}
Enforcing the initial conditions one gets
\begin{equation}
A_n = \frac{\lambda_{2,n} \, \phi_{n,0}}{\lambda_{2,n}-\lambda_{1,n}}
\; , \qquad B_n= - \frac{\lambda_{1,n}}{\lambda_{2,n}} \, A_n \; .
\label{eq2_11}
\end{equation}
For $n>n^*$, the roots of the characteristic equation are complex conjugate:
$\lambda_{1,2}= -1/2\alpha \pm i \omega_n$,  $i=\sqrt{-1}$,
$\omega_n= \sqrt{-\Delta_n}/2 \alpha$,  and
\begin{equation}
\phi_n(\theta) = e^{-\theta/2 \alpha} \left [
C_n  \cos(\omega_n \theta) + D_n  \sin(\omega_n \theta) \right ] \, ,
\label{eq2_12}
\end{equation}
where:
\begin{equation}
C_n= \phi_{n,0} \; , \qquad D_n= \frac{C_n}{2\alpha \, \omega_n} \; .
\label{eq2_13}
\end{equation}
To sum up, the solution of the above  transport problem involving
the Cattaneo equation
can be expressed as
\begin{eqnarray}
\phi(y,\theta) & = & \sum_{n=1}^{n^*} \left [A_n \, e^{\lambda_{1,n} \theta}
+ B_n \, e^{\lambda_{2,n} \theta} \right ] \, \cos(\nu_n y) 
\label{eq2_14} \\
& = & \sum_{n=n^*+1}^\infty \left [ C_n  \cos(\omega_n \theta)
+ D_n  \sin(\omega_n \theta) \right ] \, e^{-\theta/2 \alpha} \,
\cos(\nu_n y) \; .
\nonumber
\end{eqnarray}

\begin{figure}[h!]
\begin{center}
\epsfxsize=10cm
\epsffile{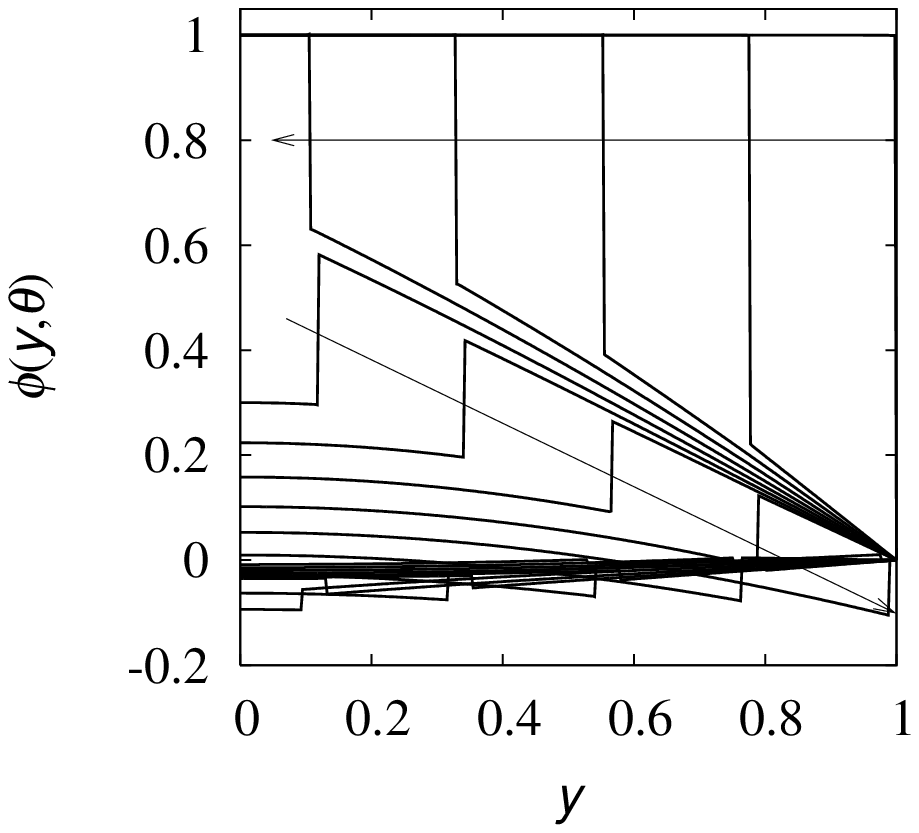} \hspace{-2.0cm} {\large (a)} \\
\epsfxsize=10cm
\epsffile{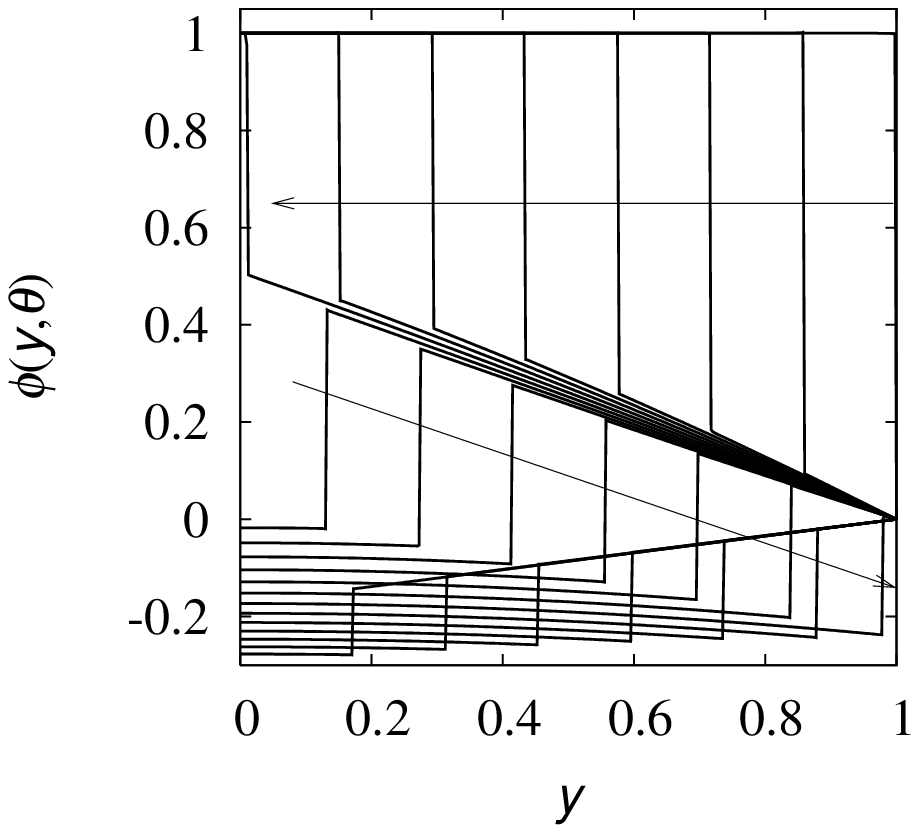} \hspace{-2.0cm} {\large (b)}
\end{center}
\caption{Concentration profiles $\phi(y,\theta)$ vs $y$
for increasing values of $\theta= n \Delta \theta$, $n=1,2,\dots$,
$\Delta \theta=0.1$ (in the direction of the arrows).
Data are obtained from eq.  (\ref{eq2_14}) truncating the
series up to $N=40000$ modes.
Panel (a) refers to $\alpha=0.2$, panel (b) to $\alpha=0.5$.}
\label{Fig1}
\end{figure}

Figure \ref{Fig1} depicts the concentration profiles for different
values
of $\theta$ along the nondimensional spatial coordinate $y$,
for $\alpha=0.2$ (panel a) and for $\alpha=0.5$ (panel b).
In both cases, it can be neatly observed that the concentration
attains negative values.

The unphysical inconsistency of the transport model is
even more evident by considering the relative released fraction
\begin{equation}
\frac{M_d(\theta)}{M_\infty}= 1-\int_0^1 \phi(y,\theta) \, d y \; .
\label{eq2_15a}
\end{equation}
In the  case under examination, since the time derivative of the flux at $y=1$
is identically vanishing for $\theta>0$, eq. (\ref{eq2_15a})
can be written in the equivalent form
\begin{equation}
\frac{M_d(\theta)}{M_\infty} = - \int_0^\theta  \left .
\frac{\partial \phi(y,\theta^\prime)}{\partial y} \right |_{y=1}
\, d \theta^\prime \, ,
\label{eq2_15}
\end{equation}
similarly to the pure Fickian case.
The quantity $M_d(\theta)/M_\infty$ represents the ratio of
solute  released up to time $\theta$ to the
overall  solute quantity initially present within the matrix.
Since there is no source, physics dictates that this quantity should
be bounded by 1, i.e., $M_d(\theta)/M_\infty \leq 1$,
$\lim_{\theta \rightarrow \infty} M_d(\theta)/M_\infty=1$.

\begin{figure}[h!]
\begin{center}
\epsfxsize=9cm
\epsffile{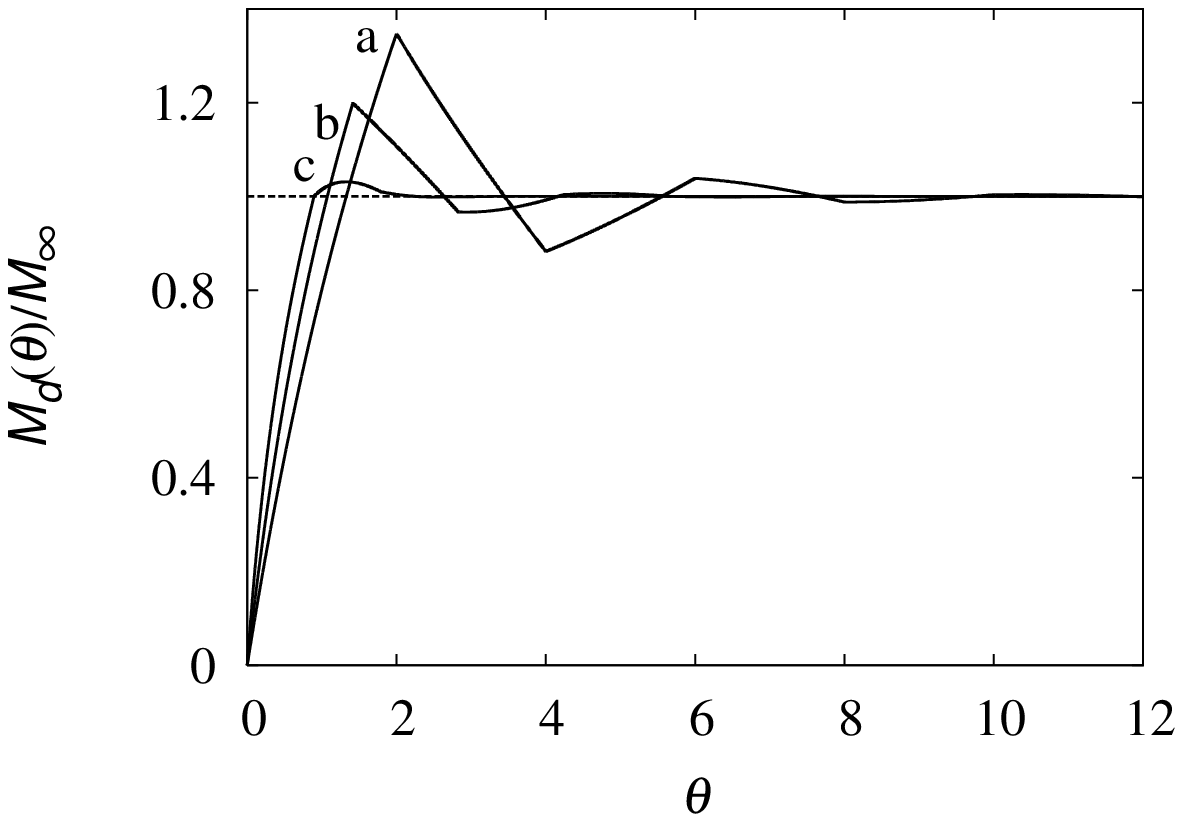}
\end{center}
\caption{$M_d(\theta)/M_\infty$ vs $\theta$ for different values
of $\alpha$. Line (a): $\alpha=1$, line (b): $\alpha=0.5$,
line (c): $\alpha=0.2$.}
\label{Fig2}
\end{figure}

Figure \ref{Fig2} depicts the time evolution of
$M_d(\theta)/M_\infty$ for several values of the dimensionless
number $\alpha$, showing the occurrence of an
overshot of this quantity reaching unphysical values
greater that 1.
This phenomenon is more pronounced  as $\alpha$ increases i.e.,
as the influence of the memory term in the constitutive
equation becomes more significant.

It should be observed that overshot dynamics in sorption experiments have
been observed in transport across polymeric matrices
where the viscoelastic response of the material
is responsible for the occurrence of memory effect in mass transport
(see \cite{polym} and reference therein, and  particularly figure 1 in this
article
based on closed-form solution of the transmission-line 
equation \cite{carslaw}). However, while in sorption experiments an overshot in the overall
sorption curve does not violate
in principle any fundamental law of nature, the
occurrence of $M_d(\theta)/M_\infty$ greater than 1 in the desorption
experiment discussed in this Section is physically inconsistent,
since it is the macroscopic manifestation of the occurrence of
negative concentration values.

\section{Stochastic dynamics, wave formulation
and boundary-condition constraints}
\label{sec3}

The paradoxes of negative concentration values and of a released
fraction greater than one at intermediate timescales provide an
interesting example of the role played by the boundary conditions,
in that the choice of admissible boundary conditions is entirely
entailed in the field equations, or better to say in the present
case, in its underlying microscopic stochastic formulation.
Different classes of boundary conditions for the Cattaneo
equation defined on an interval  have been addressed in \cite{weiss1,bound1,bound2}. The focus of the present analysis is however slightly different
with respect to these papers, as it is strictly oriented on the
consistency of the boundary conditions in order to preserve the positivity of
the solutions.

The result obtained in the previous Section could  sound as a paradox,
by considering thta the Cattaneo equation on the one-dimensional line admits
a probabilistic interpretation in term of a stochastic process
driven by dichotomous Poisson noise \cite{kac,weiss2,kolesnik2,haba}.

Consider the stochastic equation
\begin{equation}
d x(t)= c \, (-1)^{\chi(t)} \, dt \; ,
\label{eq3_1}
\end{equation}
where $c>0$ is a reference velocity, $\chi(t)$  an ordinary Poisson process
 characterized by the transition rate constant $a>0$  \cite{kac}.
 Let $X(t)$
be the stochastic process associated with eq. (\ref{eq3_1}) and let
\begin{eqnarray}
p^+(x,t)\, dx &=& \mbox{Prob}\{ X(t) \in (x,x+dx) , (-1)^{\chi(t)}=1\}
\nonumber \\
p^-(x,t)\, dx &=& \mbox{Prob}\{ X(t) \in (x,x+dx) , (-1)^{\chi(t)}=-1\} \; ,
\label{eq3_2}
\end{eqnarray}
the probabilities of the occurence of a value of $X(t)$
in the interval ($x,x+dx)$,  and of the stochastic
dichotomic
perturbation $(-1)^{\chi(t)}=\pm 1$, respectively, at time $t$.
The two partial probability densities  $p^\pm(x,t)$
specify completely  the statistical
characterization of the process (\ref{eq3_1}) \cite{kac}. 
From eqs. (\ref{eq3_1})-(\ref{eq3_2}), it follows that the evolution
 equations
for the two quantities $p^{\pm}(x,t)$ read
\begin{eqnarray}
\frac{\partial p^+(x,t)}{\partial t} & = & - c  \, \frac{\partial p^+(x,t)}{\partial x} - a p^+(x,t) + a p^-(x,t) \nonumber \\
\frac{\partial p^-(x,t)}{\partial t} & = &  c \, \frac{\partial p^-(x,t)}{\partial x} + a p^+(x,t) - a p^-(x,t) \; ,
\label{eq3_3}
\end{eqnarray}
where $a$ is the  rate constant of the Poisson process $\chi(t)$.
By defining the two quantities,
\begin{equation}
p(x,t) = p^+(x,t)+p^-(x,t) \, , \qquad J(x,t)= c \, [p^+(x,t)-p^-(x,t)] \,
\label{eq3_4}
\end{equation}
it follows  that $p(x,t)$ satisfies the conservation equation
\begin{equation}
\frac{\partial p(x,t)}{\partial t}= -\frac{\partial J(x,t)}{\partial x} \, ,
\label{eq3_5}
\end{equation}
while $J(x,t)$ fulfills the relation
\begin{equation}
\frac{1}{2 a} \frac{\partial J(x,t)}{\partial t} + J(x,t) =
- \frac{c^2}{2 a} \, \frac{\partial p(x,t)}{\partial x} \; .
\label{eq3_6}
\end{equation}
The two scalar fields $p(x,t)$ and $J(x,t)$ correspond respectively to
the overall probability density function and its flux associated with
the stochastic dynamics (\ref{eq3_1}), and eqs. (\ref{eq3_5})-(\ref{eq3_6})
return the Cattaneo transport equation by identifying
$\tau_c= 1/2 a$ and $k=c^2/2 a$.
It follows from the definition, that the quantity $p(x,t)$ should be
strictly positive due to its probabilistic interpretation.
Let us refer to $p^+(x,t)$ and $p^-(x,t)$ as the partial probability
densities (or waves), and to the equations (\ref{eq3_3}) as the {\em partial
probability model} (acronym PPM).

Starting from the decomposition
into partial probabilities, it is possible to develop a theory for
the boundary conditions. The partial probabilities are the
fundamental physical quantities in a transport process on the line
driven by dichotomous noise, while $p(x,t)$ and $J(x,t)$ should be viewed
as derived quantities. Consequently, the boundary conditions
should be properly expressed in terms of $p^+(x,t)$ and $p^-(x,t)$ to
ensure positivity. This concept can be expressed 
 in an alternative way: physically
admissible boundary conditions for the hyperbolic  transport problem on 
the interval $[0,L]$ are those
that preserve the positivity of the partial waves $p^+(x,t)$, $p^-(x,t)$, 
starting
from non-negative (initial and boundary)  values of these two quantities.
The application of this approach to  the setting of the boundary conditions
explains the apparent paradox addressed in Section \ref{sec2}.

Consider the boundary conditions for the problem addressed in Section
\ref{sec2}, namely $J|_{x=0}=0$ and eq. (\ref{eq2_5}), i.e.,
in the present notation $p(1,t)=0$. The impermeability
condition at
$x=0$ implies for the partial probabilities
\begin{equation}
\left . p^+(x,t) \right |_{x=0} = \left . p^-(x,t) \right |_{x=0} \; ,
\label{eq3_7}
\end{equation}
which is a fully admissible boundary condition. Eq. (\ref{eq3_7})
implies that the forward propagating wave ($p^+$) equals
the incoming regressive wave $p^-$ at $x=0$.
Next, consider the other boundary condition eq. (\ref{eq2_5}).
It means that $\left . [p^+(x,t) + p^-(x,t)] \right |_{x=L}=0$, i.e.,
\begin{equation}
\left . p^{-}(x,t) \right |_{x=L} = - \left . p^+(x,t) \right |_{x=L} \;,
\label{eq3_8}
\end{equation}
implying that, if the forward probability is positive at $x=L$, the
``reflected'' backward wave should attain negative values.  Clearly,
it is
  not  an admissible
boundary condition, the consequence of which is the occurrence
of negative values for $p^+(x,t) + p^-(x,t)$.

In a similar way, it is possible to address other kinds of boundary
conditions. A paradigmatic example is given by 
radiative boundary conditions arising e.g. in
interphase transport across a boundary layer.
Below, we analyze this case.
The prototype of homogeneous radiative boundary conditions is
given by
\begin{equation}
\left . J(x,t) \right |_{x=L} = h \, \left .  p(x,t) \right |_{x=L} \; ,
\label{eq3_9}
\end{equation}
where $h$ (possessing the physical dimension of a velocity $m/s$)
is the mass-transfer coefficient across a boundary layer at $x=L$.
 Substituting the expressions for
$p$ and $J$ in terms of partial probabilities, eq. (\ref{eq3_9}) becomes
\begin{equation}
\left . p^-(x,t) \right |_{x=L} = \frac{c - h}{c + h} \, \left .
 p^+(x,t) \right |_{x=L} \; .
\label{eq3_10}
\end{equation}
The latter expression indicates that positivity is preserved provided
that the condition
\begin{equation}
c-h \geq 0 
\label{eq3_10a}
\end{equation}
is fulfilled.
This condition can be expressed in terms of the physical parameters $k$
and $\tau_c$ resulting
\begin{equation}
De_h = \frac{h^2 \tau_c}{k}   \leq 1 \; .
\label{eq3_11}
\end{equation}
where the dimensionless group $De_h$ can be referred to
as the {\em radiative Deborah number}. The radiative
Deborah number is the ratio of the characteristic ``recombination time''
$\tau_c$ (the intrinsic memory timescale) to the diffusion (conduction)
time associated with the transfer across 
the boundary layer $t_{bl}= \delta^2/k$,
where $\delta$ is the boundary-layer width. Since the boundary-layer 
width $\delta$ is implicitly defined by the relation
$h= k/\delta$, it follows that $t_{bl}=k/h^2$.

Eq. (\ref{eq3_11}) indicates that for the transport problem under
consideration, radiative boundary conditions are admissible
(i.e. they preserve positivity) provided that the Deborah number $De_h$
is less than or at most equal to 1.

\section{Revisiting the boundary conditions}
\label{sec4}

In the previous Section we have derived the more general form of
linear admissible boundary condition in the framework of the
one-dimensional PPM:
\begin{equation}
\left . p^-(y,\theta) \right |_{y=1} = \gamma \, \left . p^+(y,\theta) \right |_{y=1} \; ,
\label{eq4_1}
\end{equation}
where $\gamma \in [0,1]$ is a dimensionless parameter
depending solely  on the Deborah number $De_h$.
Enforcing eq. (\ref{eq4_1}) at $y=1$, and  the zero-flux boundary conditions
at $y=0$, it is possible to obtain physically
consistent results regarding desorption kinetics in hyperbolic transport.

\begin{figure}[h!]
\begin{center}
\epsfxsize=10cm
\epsffile{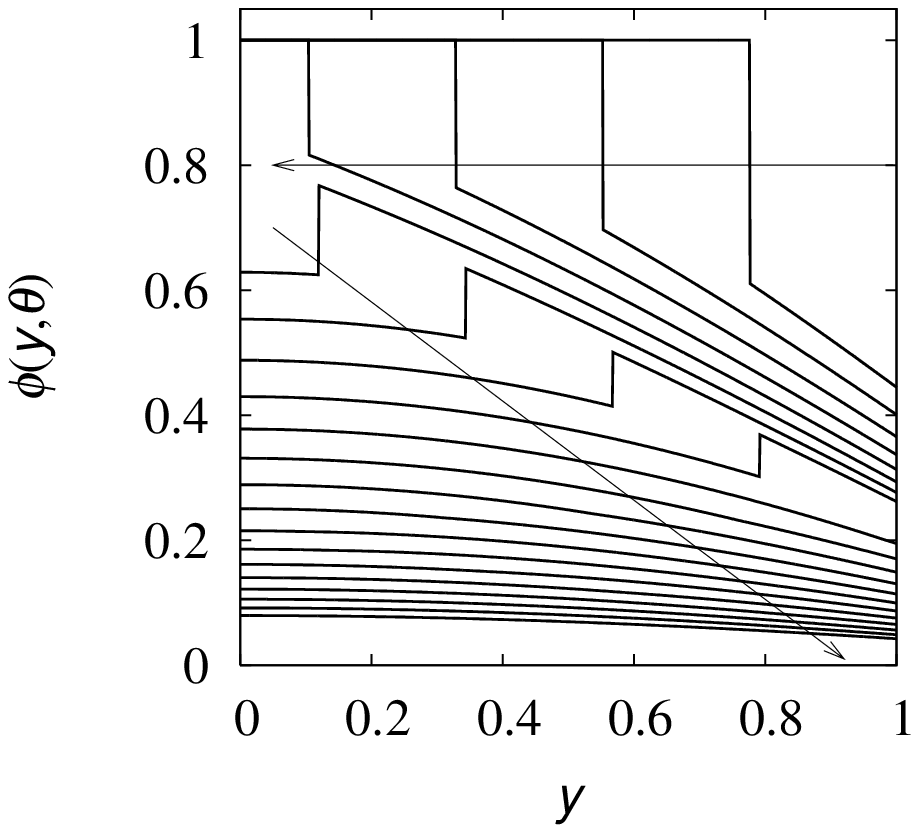} \hspace{-2.0cm} {\large (a)}\\
\epsfxsize=10cm
\epsffile{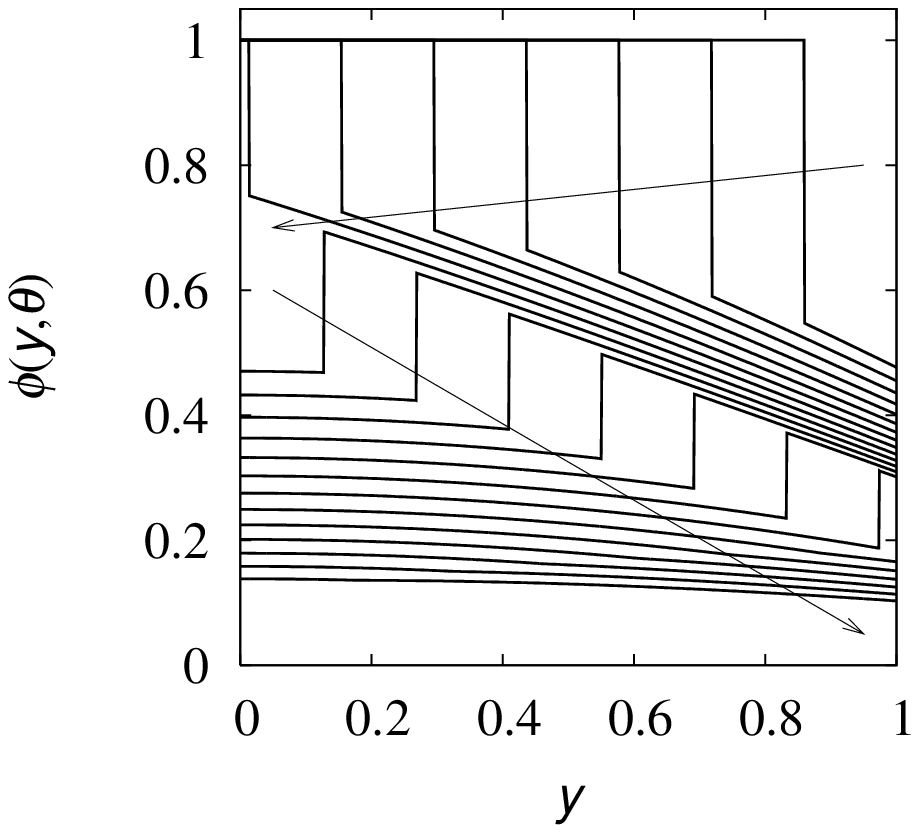} \hspace{-2.0cm} {\large (b)}
\end{center}
\caption{Concentration profiles $\phi(y,\theta)$ vs $y$
solutions of the PPM with
the boundary condition (\ref{eq4_1}) at $y=1$, $\gamma=0$
for increasing values of $\theta= n \Delta \theta$, $n=1,2,\dots$,
$\Delta \theta=0.1$ (in the direction of the arrows).
Panel (a) refers to $\alpha=0.2$, panel (b) to $\alpha=0.5$.}
\label{Fig3}
\end{figure}

Figure \ref{Fig3} depicts the overall concentration profile
$\phi(y,\theta)=p(y,\theta)=p^+(y,\theta)+p^-(y,\theta)$,
solution of the dimensionless PPM, equipped with
the boundary condition (\ref{eq4_1}), $\gamma=0$ at $y=1$
for two different values of the dimensionless parameter $\alpha$.
The initial conditions are $p^+(y,0)=p^-(y,0)=1/2$ uniformly
in $y \in (0,1)$.
As expected from the analysis of admissible boundary conditions developed in Section
\ref{sec3},
the occurrence of negative concentration values is ``cured'' by eq.
(\ref{eq4_1}).

\begin{figure}[h!]
\begin{center}
\epsfxsize=10cm
\epsffile{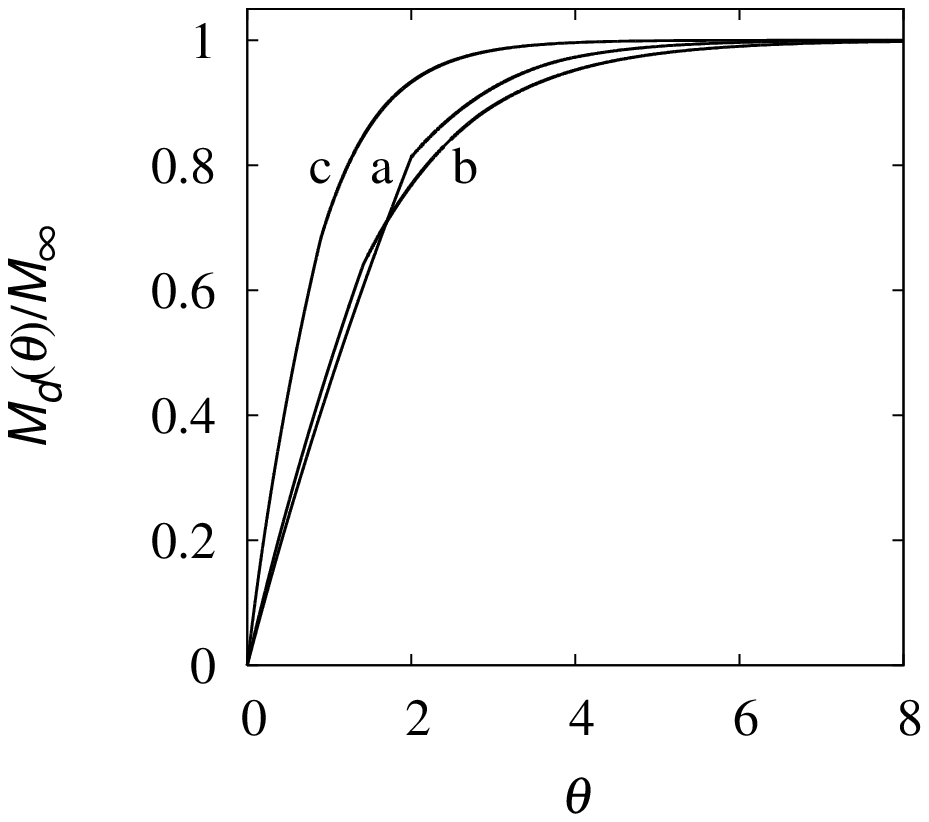} \hspace{-2.0cm} {\large (a)} \\
\epsfxsize=10cm
\epsffile{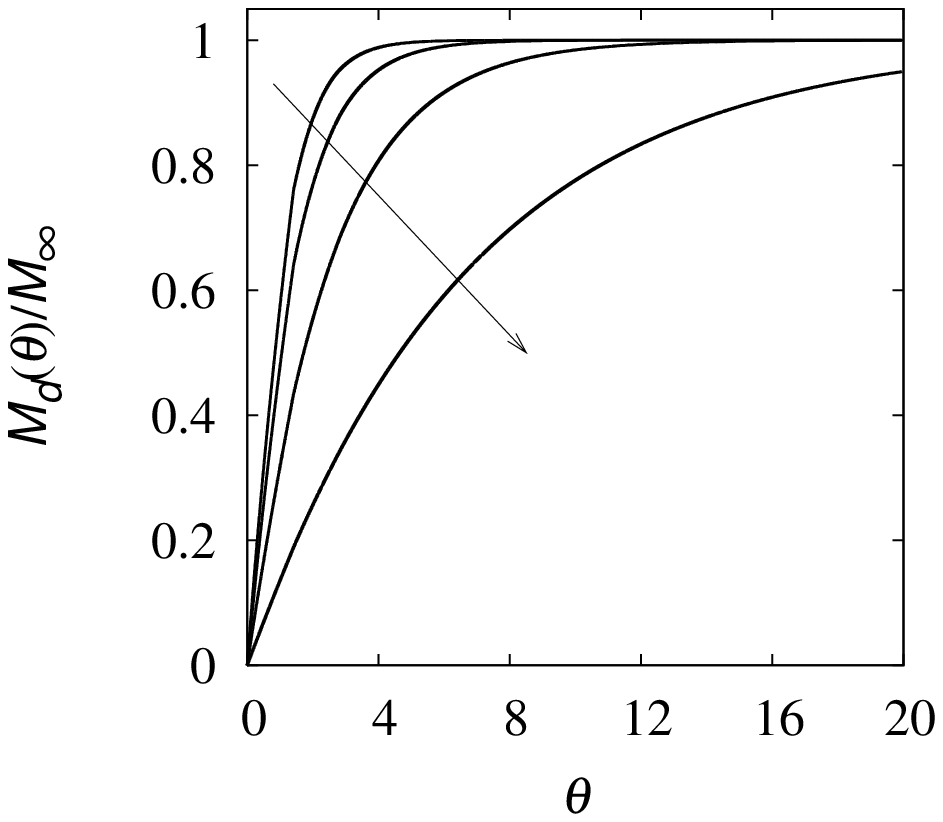} \hspace{-2.0cm} {\large (b)}
\end{center}
\caption{$M_d(\theta)/M_\infty$ vs $\theta$
 obtained from the PPM.
Panel (a): $\gamma=0$,
for different values
of $\alpha$. Line (a): $\alpha=1$, line (b): $\alpha=0.5$,
line (c): $\alpha=0.2$. Panel (b) $\alpha=0.5$, for different values of $\gamma$. The arrow
indicates increasing values of $\gamma=0,\, 0.2,\,0.5,\,0.8$.}
\label{Fig4}
\end{figure}

The overall desorption kinetics, expressed by the relative release
fraction $M_d(\theta)/M_\infty$, is depicted in figure \ref{Fig4}.
Panel (a) refers to $\gamma=0$ for different values of $\alpha$,
while panel (b) depicts the influence of the boundary parameter $\gamma$.
As expected, the release kinetics becomes progressively slower as $\gamma$
approaches $1$, since $\gamma=1$  corresponds  to the impermeability
(zero-flux) condition.
The data depicted in figure \ref{Fig4} indicate that the
unphysical overshot of $M_d(\theta)/M_\infty$ up  to values
greater than 1 cannot occur if proper boundary conditions are
enforced,
accounting for the wave-like propagation of the stochastic perturbation.

With reference to the desorption experiment analyzed throughout this work,
let
\begin{equation}
p(y,\theta)=1 + u(y,\theta) \; .
\label{eq4_2}
\end{equation}
Since $0 \leq p(y,\theta) \leq 1$, it follows that
$-1 \leq u(y,\theta) \leq 0$. In terms of partial waves
this implies
\begin{equation}
p^+(y,\theta)= \frac{1}{2} + u^+(y,\theta) \; , \qquad
p^-(y,\theta)= \frac{1}{2} + u^-(y,\theta)  \; .
\label{eq4_3}
\end{equation}
The balance equations for the partial waves $u^\pm(y,\theta)$ are
identical to eqs. (\ref{eq3_3}) replacing $p^\pm$ with $u^\pm$,
$t$ with $\theta$ and $x$ with $y$.
The initial conditions are $u^+(y,0)=u^-(y,0)=0$, and impermeability
at $y=0$ is just $u^+(0,\theta)=u^-(0,\theta)$.

More interesting is the boundary condition at $y=1$. From eqs. (\ref{eq4_1})
and (\ref{eq4_3}) one obtains
\begin{equation}
\left . u^-(y,\theta) \right |_{y=1} = \gamma \, \left . u^+(y,\theta) \right |_{y=1}- \frac{1-\gamma}{2} \; .
\label{eq4_4}
\end{equation}
This indicates that the desorption kinetics
can be solved by considering the
complementary partial waves  $u^\pm(y,\theta)$ equipped
at the transfer boundary ($y=1$) with the condition (\ref{eq4_4}).
This is a non-homogeneous expression containing a strictly negative source
term $-(1-\gamma)/2<0$, since $\gamma \in (0,1)$.

This elementary observation suggests a way to handle
boundary conditions in a sorption experiment, which 
is the complementary problem to the desorption kinetics analyzed so far.
In a sorption experiment,
solute diffuses within the solid matrix from $y=1$,
where it is in contact with a perfect reservoir
at constant unit dimensionless concentration. Assume 
a vanishing initial condition in $y \in (0,1)$.
The boundary condition at  the transfer interface $y=1$,
that  is complementary to the
desorption condition (\ref{eq4_1}) is simply
\begin{equation}
\left . u^-(y,\theta) \right |_{y=1} = \gamma \, \left . u^+(y,\theta) \right |_{y=1}+ \frac{1-\gamma}{2} \; .
\label{eq4_5}
\end{equation}
It corresponds to a change of sign in the source term entering the
desorption conditions  (\ref{eq4_4}).
Initial conditions are $u^+|_{\theta=0}=u^- |_{\theta=0}=0$, and
 at $y=0$ is enforced.

In a sorption experiment, the overall uptake kinetics is expressed
by the sorption curve
\begin{equation}
\frac{M_s(\theta)}{M_\infty} = \int_0^1 \left [ u^+(y,\theta)+ u^-(y,\theta) \right ] \, dy \; .
\label{eq4_6}
\end{equation}

\begin{figure}[h!]
\begin{center}
\epsfxsize=10cm
\epsffile{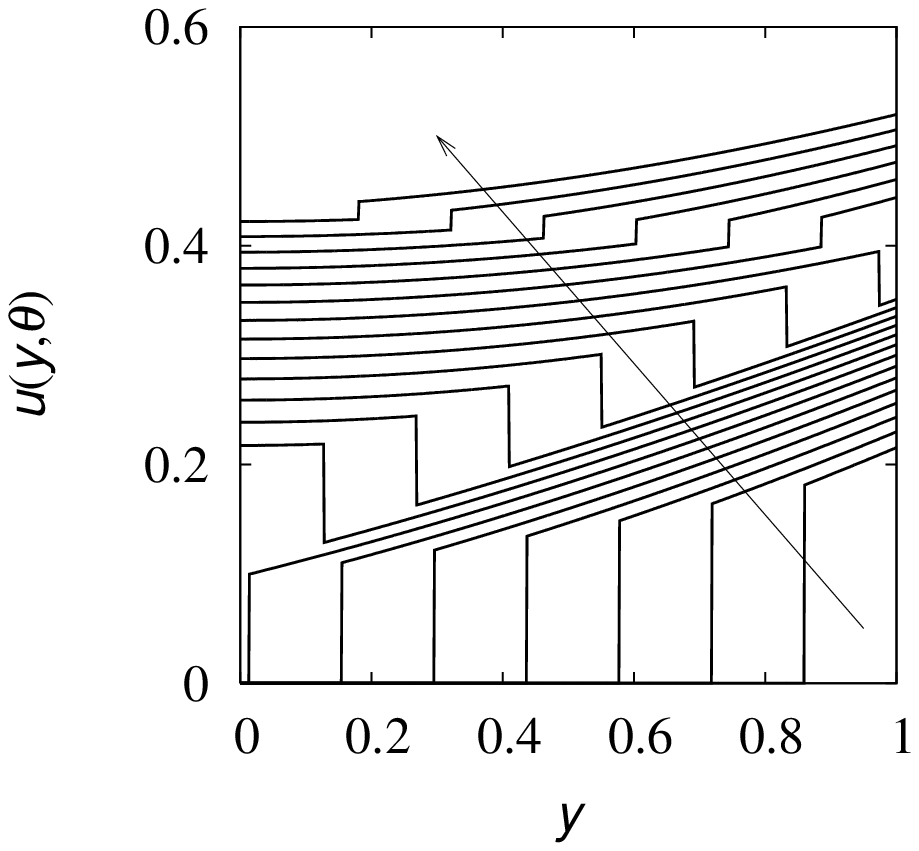} \hspace{-2.0cm} {\large (a)} \\
\epsfxsize=10cm
\epsffile{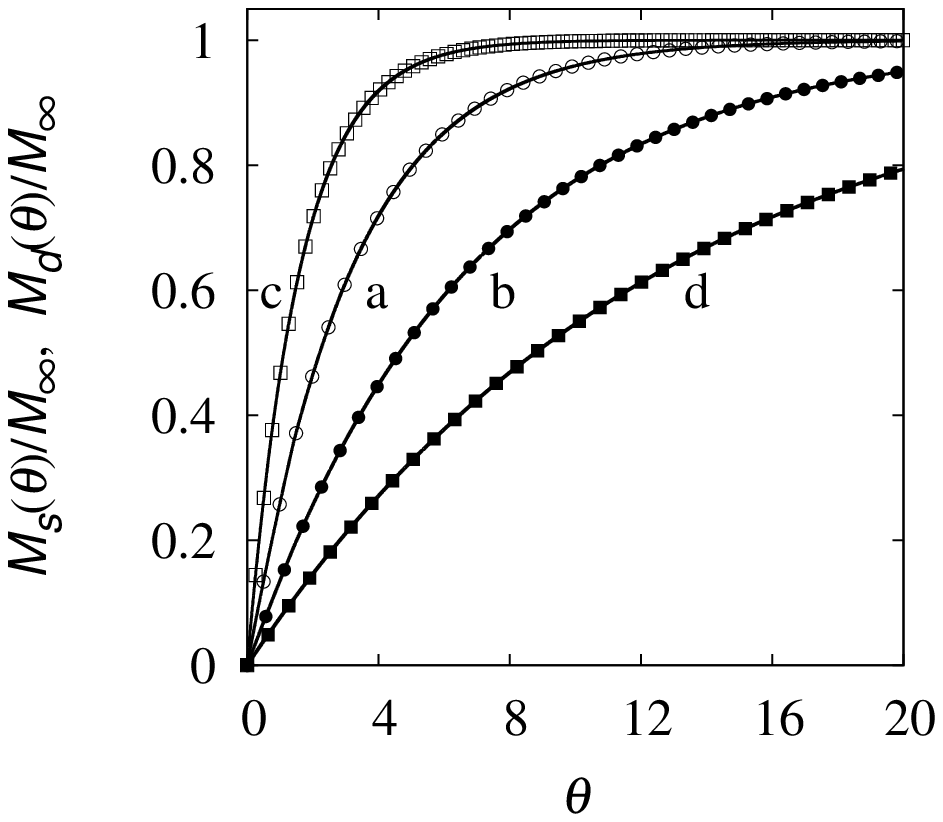} \hspace{-2.0cm} {\large (b)}
\end{center}
\caption{Panel (a): Concentration profiles in a sorption
experiment at $\tau_c=0.5$, imposing the boundary
conditions (\ref{eq4_5})  with $\gamma=0.6$, 
sampled at $\Delta \theta=0.1$ (time
increases in the direction of the arrow).  Panel
(b): Sorption curves $M_s(\theta)/M_\infty$ from eq. (\ref{eq4_6})
(solid lines)
for different values of $\tau_c$ and $\gamma$. Symbols
are the corresponding desorption curves $M_d(\theta)/M_\infty$.
 Line (a) and
($\circ$): $\tau_c=0.5$, $\gamma=0.6$, line (b) and ($\bullet$):
$\tau_c=0.5$, $\gamma=0.8$, line (c) and ($\square)$: $\tau_c=0.1$
and $\gamma=0.6$, line (d) and ($\blacksquare$):  $\tau_c=0.1$
and $\gamma=0.95$.}
\label{Fig5}
\end{figure}

Figure \ref{Fig5} (panel a) depicts the concentration profiles
for $\tau_c=0.5$, $\gamma=0.6$ at different time instants sampled at
$\Delta \theta=0.1$. Panel (b) of the same figure compares the
resulting sorption curves (solid lines) obtained
using eq. (\ref{eq4_6}) and the corresponding 
 desorption curve associated with eq. (\ref{eq4_1}). As expected,
 sorption and desorption
curves at the same values of $\tau_c$ and $\gamma$ coincide,
confirming that the nonhomogeneous boundary conditions (\ref{eq4_5})
represents in sorption experiments  the analogue of the desorption
boundary conditions (\ref{eq4_1}).

\begin{figure}[h!]
\begin{center}
\epsfxsize=10cm
\epsffile{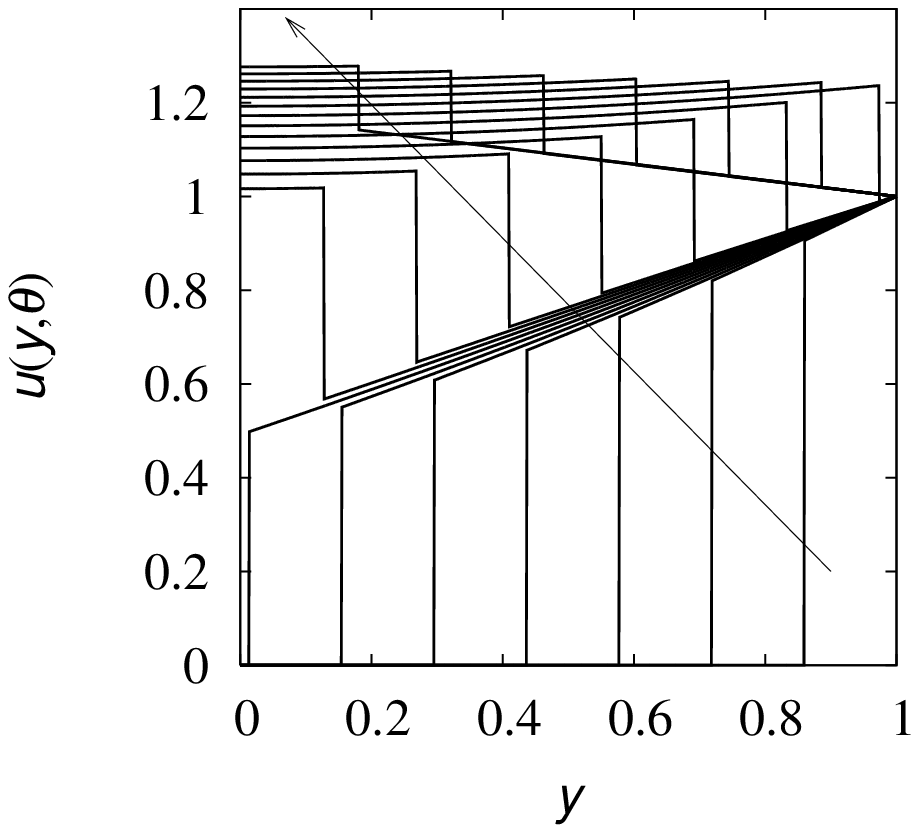} \hspace{-2.0cm} {\large (a)} \\
\epsfxsize=10cm
\epsffile{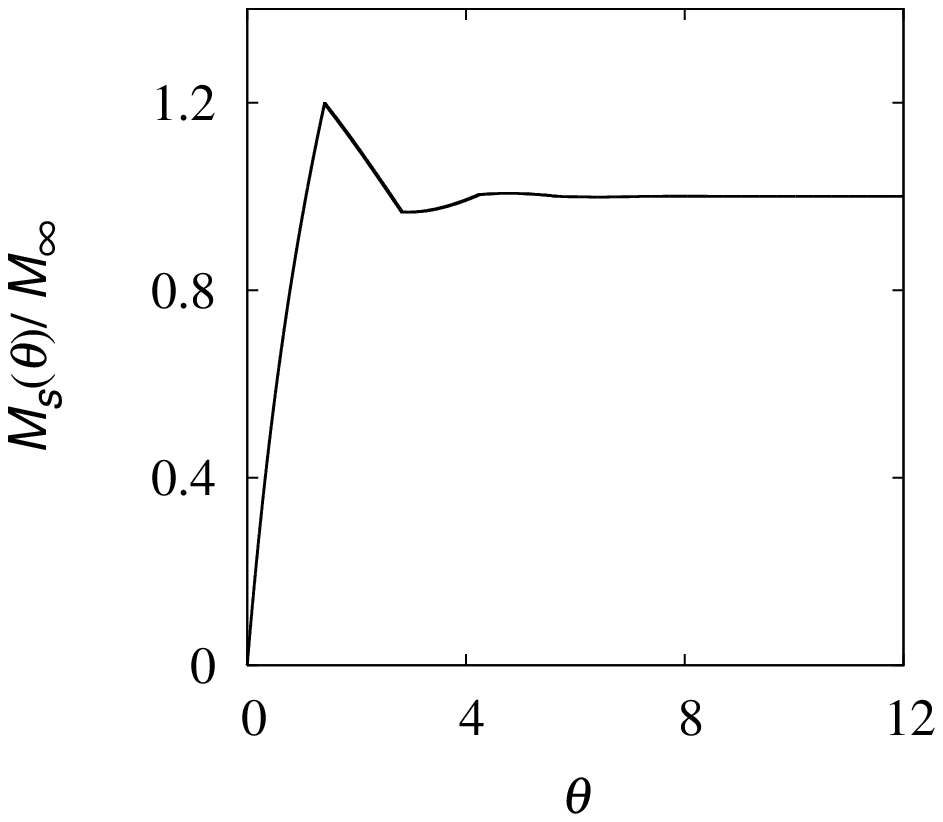}  \hspace{-2.0cm} {\large (b)}
\end{center}
\caption{Panel (a): Concentration profiles $u(y,\theta)$ in a sorption
experiment imposing the boundary condition (\ref{eq4_7}), $\tau_c=0.5$.
The curves are sampled at $\Delta \theta=0.1$ (time increases in the direction
of the arrow). Panel (b) Sorption curve $M_s(\theta)/M_\infty$
associated with the data depicted in panel (a).}
\label{Fig6}
\end{figure}

It is important to stress once again that this complementary
derives from the intrinsic decomposition of the concentration
field into partial waves. This decomposition, with
all of its implications, is the essence of hyperbolic transport.
Any attempt in modeling  boundary conditions in 
sorption dynamics starting directly
from the overall concentration field $u(y,\theta)$, thus
neglecting the underlying decomposition in partial waves $u^\pm(y,\theta)$,
leads unavoidably to physical inconsistencies. Let us
clarify this statement with a simple example. In an ideal
sorption experiment (starting from $u|_{\theta=0}=0$),
the simplest boundary condition corresponding to
the contact  with an ideal reservoir at unit concentration
would be
\begin{equation}
\left . u(y,\theta)  \right |_{y=1}=1 \; , 
\label{eq4_7}
\end{equation}
in the absence of transfer resistances.
Theuse of this equation, or of the equivalent expression,
\begin{equation}
\left . u^-(y,\theta) \right |_{y=1}= 1- \left . u^+(y,\theta) \right |_{y=1}
\label{eq4_8}
\end{equation}
ends up with an overshot dynamics for the sorption curve 
$M_s(\theta)/M_\infty$, corresponding to local concentration
values greater than 1, i.e. greater than 
the feeding concentration at the boundary $y=1$.  This is shown in figure 
\ref{Fig6} depicting the concentration profiles (panel a),
and the overall sorption curve (panel b) at $\tau_c=0.5$.

\section{Norm dynamics and dissipation}
\label{sec5}

The constraints on  the structure of the boundary conditions can be derived from
the evolution of the integral norms, enforcing dissipation.
Consider  eq. (\ref{eq3_3}) defined on an interval ${\mathcal D}$.
So far no assumption is made for ${\mathcal D}$, be it $(-\infty,\infty)$
or $[0,L]$. Multiplying the first equation (\ref{eq3_3}) by $p^+$,
the second by $p^-$, summing the resulting equations, and integrating
over ${\mathcal D}$ provides the expression
\begin{eqnarray}
\frac{1}{2} \, 
\frac{d }{d t} \int_{\mathcal D} \left [ (p^+)^2+ (p^-)^2 \right ] \, d x
& = &  - \frac{c}{2} \int_{\mathcal D} \frac{\partial}{\partial x} \left [ (p^+)^2- (p^-)^2 \right ] \, d x
\nonumber \\
  &- & a \int_{\mathcal D} \left ( p^+ - p^- \right )^2 \, d x \; .
\label{eq5_1}
\end{eqnarray}
Let us express the quantities entering eq. (\ref{eq5_1}) in terms
of concentration and flux:
\begin{eqnarray}
 (p^+)^2+ (p^-)^2 & = & \frac{1}{2}  \left ( p^+ + p^- \right )^2 
+   \frac{1}{2} \left ( p^+ - p^- \right )^2 = \frac{1}{2} \, p^2 + \frac{1}{2 \, c^2} \, J^2
\nonumber \\
 (p^+)^2- (p^-)^2 & = & (p^++p^-) (p^+-p^-) = \frac{1}{c}  \, p \, J  \; .
\label{eq5_2}
\end{eqnarray}
Two cases can be considered: either ${\mathcal D}=(-\infty,\infty)$
or ${\mathcal D}=[0,L]$.

On the real line, i.e. ${\mathcal D}=(-\infty,\infty)$, regularity
conditions at infinity apply, and consequently the
first integral at the r.h.s. of eq. (\ref{eq5_1}) vanishes
identically.

Letting $|| f||_{L^2}^2(t)= \int_{\mathcal D} f^2(x,t) \, d x$ be  the
square $L^2$-norm of a square-summable  real-valued
function $f(x,t)$ in ${\mathcal D}$,
eq. (\ref{eq5_1}) can be rewritten as:
\begin{equation}
\frac{d}{d t} \left ( ||p||_{L^2}^2(t) + \frac{1}{c^2} \,
||J||_{L^2}^2(t) \right ) = - \frac{2}{k} \, ||J||_{L^2}^2(t) \; ,
\label{eq5_3}
\end{equation}
which expresses the dissipation properties of hyperbolic ``diffusion'' i.e.
of the stochastic motion driven by pure dichotomous Poisson noise.
In the limit $c \rightarrow \infty$, $J \rightarrow -k \partial p/\partial x$,
 eq. (\ref{eq5_3}) reduces to the classical norm condition,
$d ||p||_{L^2}^2(t)/dt = - 2 k \, ||\partial p/\partial x||_{L^2}^2(t)$,
characterizing Fickian diffusion.

Next, consider a bounded domain ${\mathcal D}=[0,L]$
and let us suppose (as in the previous Sections) that
at $x=0$ zero-flux conditions are enforced.
Eq. (\ref{eq5_1}) now becomes
\begin{equation}
\frac{d}{d t} \left ( ||p||_{L^2}^2(t) + \frac{1}{c^2} \,
||J||_{L^2}^2(t) \right ) = -I_{\rm boundary}- \frac{2}{k} \, ||J||_{L^2}^2(t) \; ,
\label{eq5_4}
\end{equation}
where the boundary term $I_{\rm boundary}$ equals
\begin{eqnarray}
I_{\rm boundary} & = & 2 \, p(L,t) \, J(L,t) \nonumber \\
& = & 2 c \, \left [p^+(L,t)+p^-(L,t) \right ] 
 \left [p^+(L,t)-p^-(L,t) \right ] \; .
\label{eq5_5}
\end{eqnarray}
Irreversibility, or equivalently  the dissipative nature of the process,
imply that the r.h.s. of eq. (\ref{eq5_4}) should be strictly non-positive.
Consider at $x=L$ homogeneous boundary conditions for the partial
waves, such as in eq. (\ref{eq4_1}). A sufficient (not necessary)
condition is that the constant $\gamma$ entering eq. (\ref{eq4_1})
satisfies the inequality $\gamma \leq 1$, which is the
condition obtained in Section \ref{sec4} from positivity arguments.

\section{Biased hyperbolic transport and boundary-layer polarization}
\label{sec6}

In this Section we extend the analysis of boundary conditions
in  hyperbolic  models to 
 biased transport. Specifically, let $v(x)$
be a deterministic velocity field, and consider the stochastic
dynamics:
\begin{equation}
d x(t)= v(x(t)) d t + c \, (-1)^{\chi(t)} \, d t \; ,
\label{eq6_1}
\end{equation}
defined in the interval $x \in [0,L]$. The PPM  that specifies the
statistical properties of eq. (\ref{eq6_1}) is expressed by the
system of two balance equations
\begin{eqnarray}
\frac{\partial p^+(x,t)}{\partial t} & = & -
\frac{\partial \left [ v(x) \, p^+(x,t) \right ]}{\partial x}
- c \, \frac{\partial p^+(x,t)}{\partial x} -a \, p^{+}(x,t)+a 
\, p^-(x,t) \nonumber \\
\frac{\partial p^-(x,t)}{\partial t} & = & -
\frac{\partial \left [ v(x) \, p^-(x,t) \right ]}{\partial x}
+ c \, \frac{\partial p^-(x,t)}{\partial x} +a \, p^{+}(x,t)-a 
\, p^-(x,t)  \nonumber \\
& & \label{eq6_2}
\end{eqnarray}
where $p^{\pm}(x,t)$ are defined as in eq. (\ref{eq3_2}).

Let us study the evolution of the overall concentration field $p(x,t)$
and of its partial waves in a closed system, supposing that $v(0)=0$,
while $v(L)>0$. For instance, consider the simple
model for the biasing field
\begin{equation}
v(x)= - V_L \sin \left ( \frac{ 3 \pi x}{ 2 L} \right ) \; ,
\label{eq6_3}
\end{equation}
where $V_L>0$. Since $v(L)>0$, there is a nonvanishing, outwardly
directed, convective flux at $x=L$.  The closedness of the
system
 implies that
the net flux vanishes at $x=0,L$. Since $v(0)=0$,
reflecting boundary conditions for the partial waves
apply at $x=0$, i.e. eq. (\ref{eq3_7}),  as in the previous examples.

More interesting is the analysis at $x=L$, where the vanishing of the
net flux dictates
\begin{equation}
v(L) \, p(L,t) + J(L,t) =0 \; ,
\label{eq6_4}
\end{equation}
where $J(L,t)$ is the ``diffusive'' (non-Fickian) flux associated
with the stochastic Poisson perturbation in eq. (\ref{eq6_1}).
In terms of partial waves this implies
\begin{equation}
v(L) \left [ p^+(L,t) + p^-(L,y) \right ] + c \, \left  [ p^+(L,t) - p^-(L,y) \right ]  = 0 \; ,
\label{eq6_5}
\end{equation}
meaning  that the ``diffusive'' flux, proportional to $p^+-p^-$,
should  counterbalance the convective one, proportional to $p^++p^-$, creating
a concentration boundary layer ({\em polarization}) near $x=L$.
Since $v(L)=V_L$, the latter condition becomes
\begin{equation}
(c - V_L) \, p^-(L,t) = (c+ V_L) \, p^+(L,t) \; .
\label{eq6_6}
\end{equation}
Eq. (\ref{eq6_6}) applies, without returning negative
values of $p^-(L,t)$,  if the condition
\begin{equation}
c > V_L \; ,
\label{eq6_7}
\end{equation}
is fulfilled. The case $c<V_L$ provides negative values of $p^-(L,t)$,
while if $c \rightarrow V_L$  the backward wave at the boundary
diverges, $p^-(L,t) \rightarrow \infty$.

Therefore,  boundary layer polarization
can be properly defined in hyperbolic transport models 
 if eq. (\ref{eq6_7}) is
fulfilled, i.e., the intensity of the stochastic velocity perturbation, namely $c$, is strictly greater than the velocity $V_L$ at the boundary.
In this case, the proper boundary condition for the
partial waves at $x=L$ is homogeneous and equal to
\begin{equation}
p^-(L,t) = \Gamma \, p^+(L,t) \; , \qquad
\Gamma= \frac{c+ V_L}{c-V_L} \; , \qquad 1 \leq \Gamma < \infty 
\; .
\label{eq6_8}
\end{equation}

\begin{figure}[h!]
\begin{center}
\epsfxsize=10cm
\epsffile{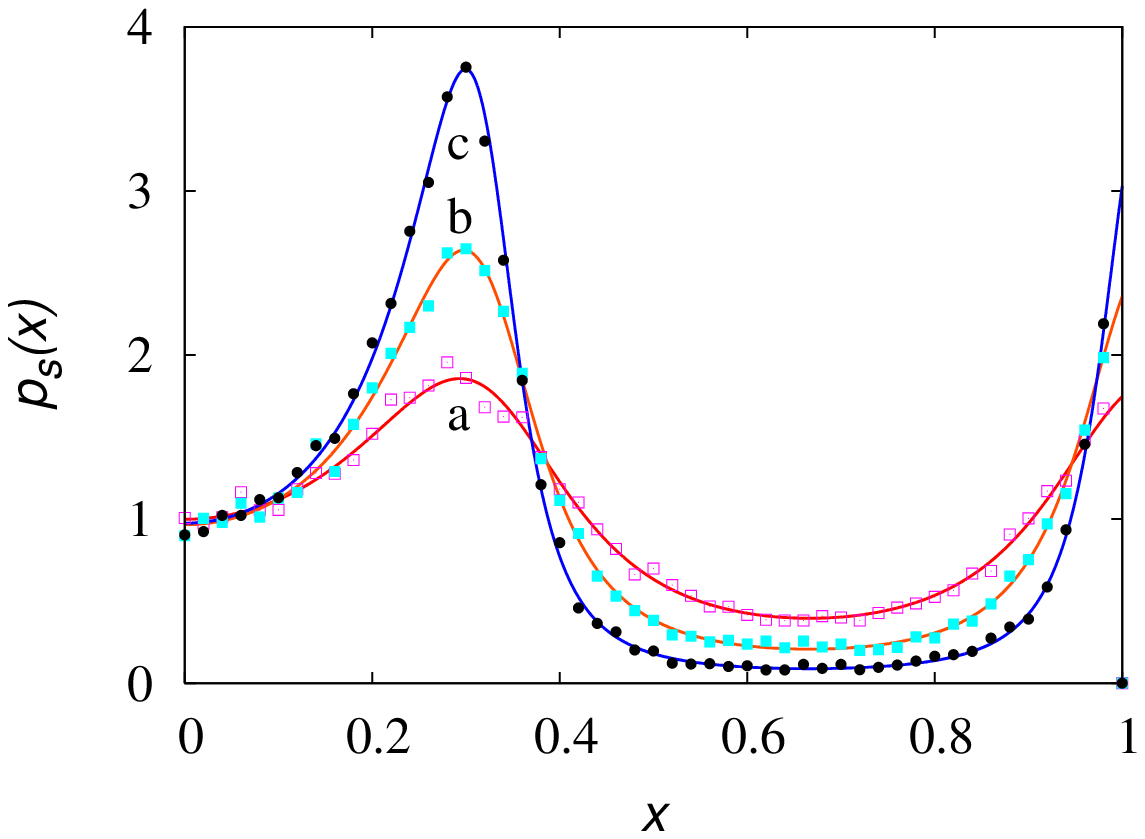} \hspace{-2.0cm} {\large (a)}\\
\epsfxsize=10cm
\epsffile{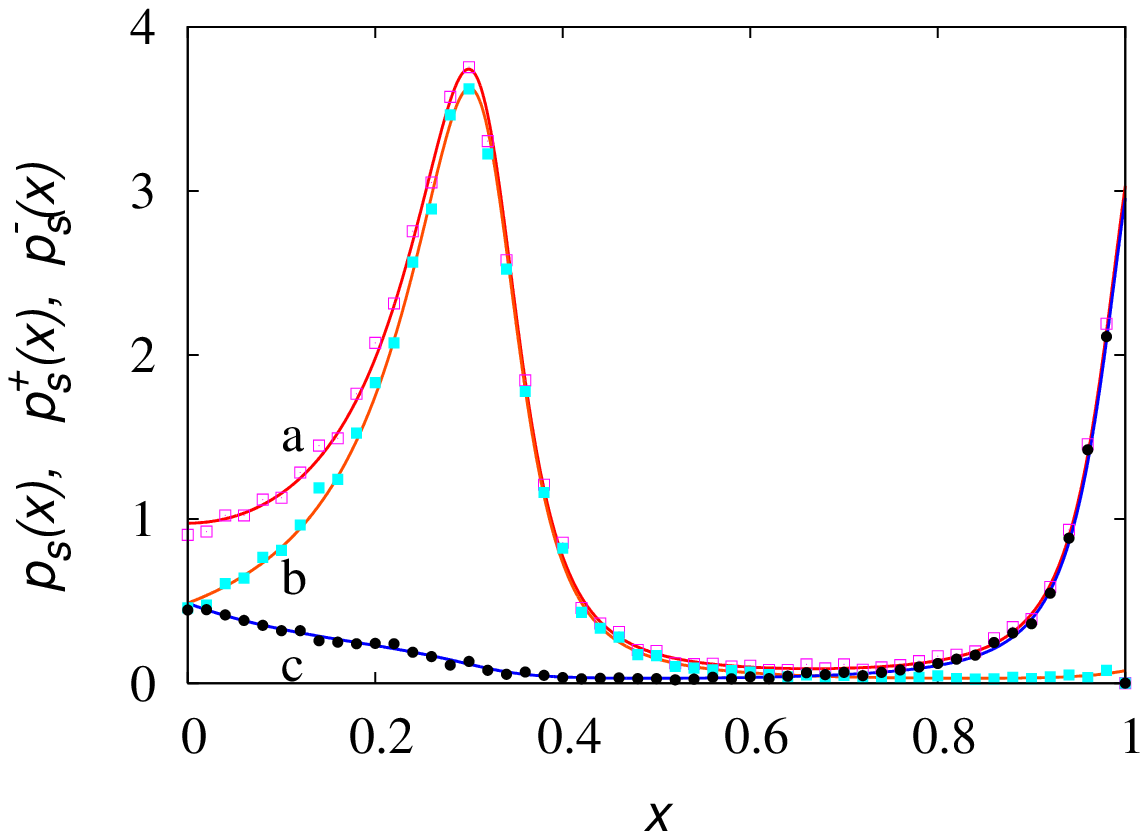}  \hspace{-2.0cm} {\large (b)}
\end{center}
\caption{Panel (a): Unique-steady state probability density function
$p_s(x)$ vs $x$ for $\tau_c=0.5$, $k=1$, at different values of $\nu$.
Solid lines are the results of the partial wave model (\ref{eq6_2}), symbols
represent stochastic simulations for an  ensemble of particles evolving
according to eq. (\ref{eq6_1}). Line (a) and ($\square$): $\nu=0.8$,
line (b) and ($\blacksquare$): $\nu=0.9$, line (c) and ($\bullet$): $\nu=0.95$.
Panel (b): Steady-state probability density function $p_s(x)$ vs $x$ (line (a)
and symbols ($\square$)),
and associated steady-state partial waves $p_s^+(x)$ (line (b) and symbols ($\blacksquare$)) and $p_x^-(x)$ (line (c) and symbols ($\bullet$))
for $\tau_c=0.5$, $k=1$, $\nu=0.95$.}
\label{Fig7}
\end{figure}

Let us consider a numerical example which highlights better the
meaning and the nature of condition (\ref{eq6_7}).
We assume for $v(x)$  the expression  (\ref{eq6_3})
and let $L=1$. Figure \ref{Fig7} (panel a) depicts the behavior of the
steady-state distribution  $p_s(x)=p_s^+(x)+p_s^-(x)$ 
associated with eq. (\ref{eq6_2}),
obtained from the asymptotic  solution of PPM eq. (\ref{eq6_2}) 
for  $t \rightarrow \infty$.
The values of the parameters are $\tau_c=0.5$, and $k=1$, and we set 
 $V_L= \nu \, c$ with $\nu<1$. The steady-state distributions 
refer to several values
of $\nu= V_L/c$. Given $\tau_c$ and $k$,
the parameters $c$, $a$ specifying the stochastic perturbation
are determined by the relations $a=1/2 \tau_c$, 
$c=\sqrt{2 k a}= \sqrt{k/\tau_c}$. The numerical results of the
PPM eq. (\ref{eq6_2}), represented by solid lines in figure 
\ref{Fig7}, are
compared  with  stochastic simulations 
by considering an ensemble of $N=2 \times 10^5$
particles following the stochastic micro-dynamics (\ref{eq6_1}).
The initial positions of the particles are uniformly distributed
in the interval $[0,1]$. Reflective boundary conditions
are assumed at $x=0$ and $x=1$. Whenever a collision at the boundaries
$x=0$ or $x=1$ occurs, say at $t=t^*$, the stochastic process
is updated. This means that at $t=t^*$, the factor $(-1)^{\chi(t)}$
changes sign, namely, $(-1)^{\chi(t^*_+)}=-(-1)^{\chi(t^*_-)}$,
where $t^*_\pm= \lim_{\varepsilon \rightarrow 0} t^*\pm\varepsilon$.
As can be observed the agreement between PPM and stochastic
simulations is excellent. The hyperbolic transport model
provides the occurrence of a polarization layer near $x=1$,
that becomes more pronounced as $\nu$ tends to $1$.

The maximum of the steady-state distribution occurs
at some intermediate abscissa $x^*$, corresponding to the stable node of the
forward wave, i.e., $v(x^*)+c=0$, and not
at $x=0$ which represents the stable equilibrium point of the
deterministic dynamics.

Keeping track of the values of $(-1)^{\chi(t)}$, it
is possible to extract from stochastic simulation data the
distribution of the two-partial waves $p_s^+(x)$ and
$p_s^-(x)$, using their
definitions  (\ref{eq3_1}) as  joint
probabilities with respect to the dichotomous outcome
of $(-1)^{\chi(t)}$. The comparison
between PPM and stochastic simulations
for the partial waves is depicted in figure \ref{Fig7}
(panel b), revealing also in this case an excellent agreement,
limited solely by the relatively small size, $N=2 \times 10^5$,
of the particle ensemble considered in the statistics.

The comparison of PPM simulation and stochastic data reveals
an interesting property of the  condition (\ref{eq6_7}). In point of fact,
there  are neither physical limitations nor 
fundamental consistency constraints in simulating an ensemble of
particles moving according to eq. (\ref{eq6_1}) beyond
the limit imposed by eq. (\ref{eq6_7}), i.e., for $V_L>c$, imposing
reflective conditions at the boundaries.

\begin{figure}[h!]
\begin{center}
\epsfxsize=10cm
\epsffile{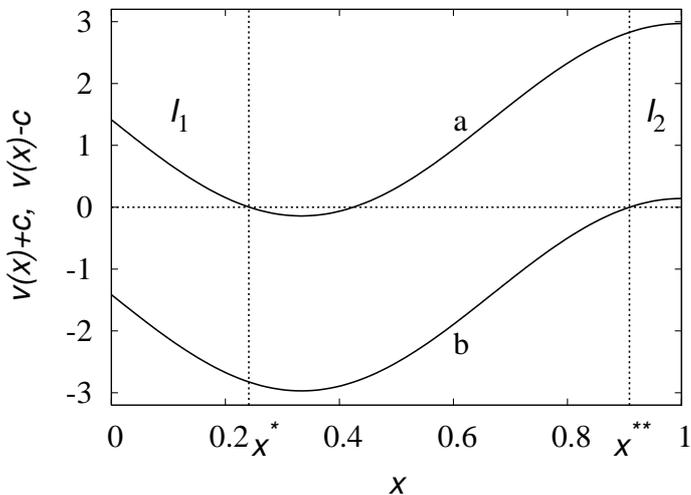}
\end{center}
\caption{$v(x)+c$ (line a) and $v(x)-c$ (line b) for $\nu=V_L/c=1.1$.
Dotted vertical lines mark the two characteristic points $x^*$,
$x^{**}$ representing the endpoints of the two invariant intervals
$I_1=[0,x^*]$ and $I_2=[x^{**},1]$.}
\label{Fig8}
\end{figure}

While eq. (\ref{eq3_10a}) stems from physical principles
(non-negativity of probabilities, or equivalently dissipativity
as developed in Section \ref{sec5}), this is not the case of
eq. (\ref{eq6_7}). The
origin of the critical condition $c=V_L$ is
not of basic physical nature, but possesses
a dynamic explanation. Namely, it is essentially a bifurcation
point of the partial-wave model, associated with the
ergodicity-breaking \cite{ergodicity_breaking}, corresponding to
the birth of multiple invariant measures.
For $c > V_L$ there is a unique invariant stationary
probability density function $p_s(x)$. Conversely,
for $c<V_L$, the stochastic dynamics (\ref{eq6_1})
is characterized by the occurrence of two
disjoint invariant intervals $I_1=[0,x^{*}]$ and $I_2=[x^{**},1]$,
where $x^*$ is the first root of the
equation $v(x^*)+c=0$ and $x^{**}$ the root of $v(x^{**})-c=0$ (see figure \ref{Fig8}).
Consequently there exist two
distinct steady-state distributions $p_{s,h}(x)$, $h=1,2$
 localized within the two intervals $I_h$.

\begin{figure}[h!]
\begin{center}
\epsfxsize=7cm
\epsffile{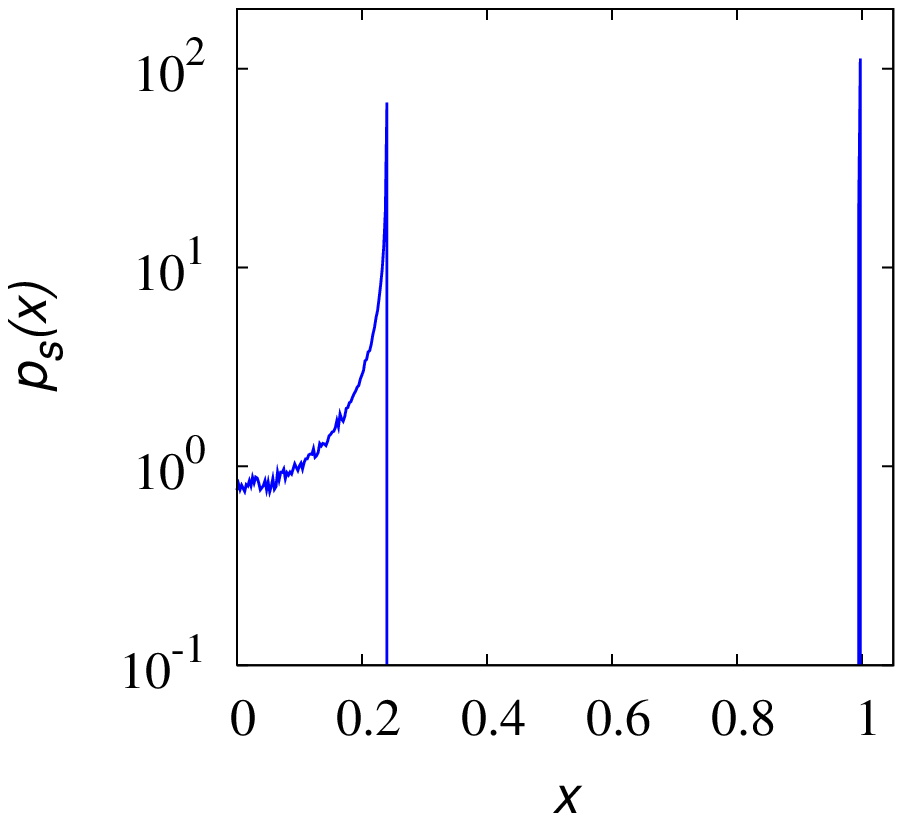} \hspace{-2.0cm} {\large (a)}\\
\epsfxsize=7cm
\epsffile{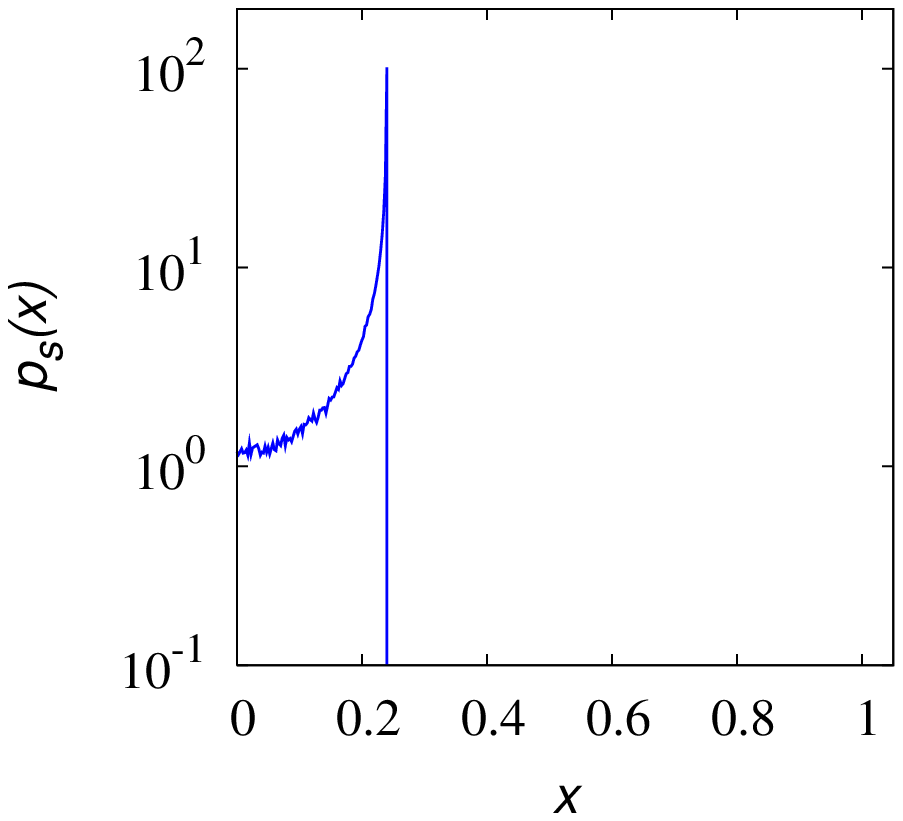}  \hspace{-2.0cm} {\large (b)}\\
\epsfxsize=7cm
\epsffile{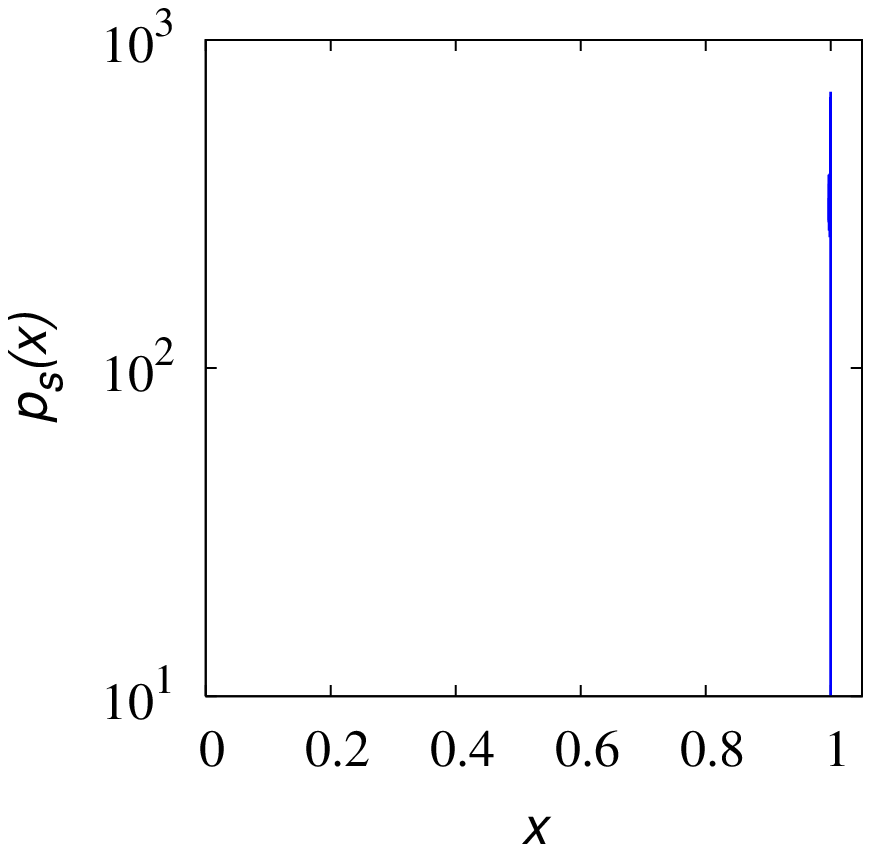}  \hspace{-2.0cm} {\large (c)}\\
\end{center}
\caption{Steady-state distributions for $\tau_c=0.5$, $k=1$, $\nu=1.1$ (i.e.,
violating condition (\ref{eq6_7})) obtained from stochastic simulation
of eq. (\ref{eq6_1}) for different distributions of 
initial conditions. Panel (a): uniform distribution in $[0,1]$,
panel (b): uniform distribution in $[0,0.3]$, panel (c): uniform distribution
in $[  0.92,1]$.}
\label{Fig9}
\end{figure}

This phenomenon can be easily verified via stochastic 
simulations starting from different initial
distributions, as depicted in figure \ref{Fig9} panels (a)-(c)
for $\nu=1.1$. In this case, $x^* \simeq 0.2412$ and  $x^{**}\simeq 0.9087$.

Panel (b) depicts the steady-state  distributions
obtained by considering an initial particle ensemble
uniformly distributed in $[0,0.3] \subset I_1$, returning
an absolutely continuous invariant distribution
localized near $x^*$.
Conversely for an initial ensemble contained in $I_2$ (depicted
in panel (c)), an impulsive invariant distribution $p_{s,2}(x)$
localized at $x=1$ is obtained.
The simulation starting from an
 uniform distribution throughout $[0,1]$, depicted in panel (a),
provides a mixed steady-state distribution $p_s(x)= \alpha p_{s,1}(x) +
(1-\alpha) p_{s,2}(x)$, $0< \alpha < 1$, that is a convex
combination of the two $p_{s,h}(x)$.

Although a thorough analysis of ergodicity-breaking bifurcations is developed
elsewhere \cite{ergodicity_breaking}, since it goes beyond the
analysis of boundary conditions representing the focus of the
present work,  it is interesting to analyze the boundary conditions
that apply to the partial-wave model  for $V_L>c$ in order to reproduce
the results obtained from stochastic simulations.
Also in the analysis  of this issue, the answer stems from
a  wave-oriented interpretation of hyperbolic
transport.

With the aid of figure \ref{Fig8}, a wave-dynamic interpretation of
the bifurcation occurring at $V_L=c$ can be achieved. From their
definition and dynamics (\ref{eq6_2}), the partial probability 
$p^+(x,t)$ and $p^-(x,t)$  corresponds to waves 
that propagate with velocities
$v^+(x)=v(x)+c$ and $v^-(x)=v(x)-c$, respectively.
For $V_L<c$,  in the neighbourhood of $x=1$, $v^+(x)>0$ and
$v^-(x)<0$, thus representing properly
a progressive and a regressive wave, respectively.

 In this
case, the balance at $x=1$ of the overall flux permits
to derive the condition  (\ref{eq6_8}) for the ``reflection
coefficient'' $\Gamma$, providing a regressive wave
that  balances  exactly the outward probability flux
carried out by the progressive wave.

The bifurcation point $V_L=c$ marks
the transition  for $p^-(x,t)$ from a regressive wave ($v^-(x)<0$) to
a progressive wave ($v^-(x)>0$), near $x=1$.
In the latter  case, impermeability at $x=1$ cannot be
longer enforced as relation connecting  $p^-(1,t)$
to $p^+(1,t)$, since $p^-(1,t)$ would be negative.
It  is rather  a consequence
of the  implicit assumption assumed  in stochastic simulations
of an arbitrarily large (and in principle infinite) potential
barrier at $x=1$ originating the reflective condition for
eq. (\ref{eq6_1}) at $x=1$.

\begin{figure}[h!]
\begin{center}
\epsfxsize=10cm
\epsffile{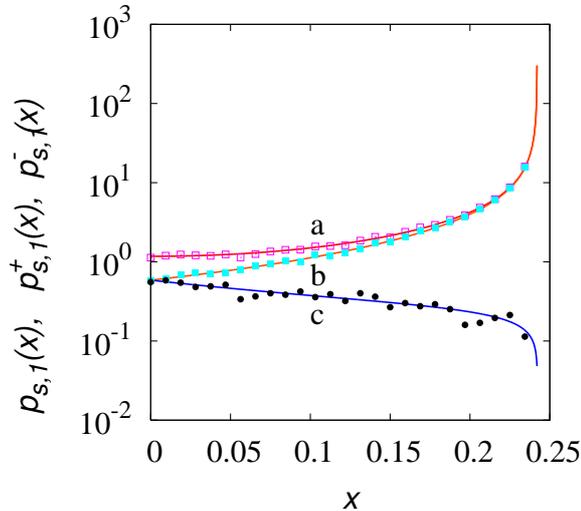}
\end{center}
\caption{Steady-state distributions $p_{s,1}(x)$,  $p_{s,1}^\pm(x)$
restricted to $I_1$ for $\tau_c=0.5$, $k=1$, $\nu=1,1$.
Solid lines represent the result of the partial wave model
enforcing condition (\ref{eq6_10}), symbols  correspond
the stochastic simulation data. Line (a) and ($\square$): $p_{s,1}(x)$,
 line (b) and ($\blacksquare$): $p_{s,1}^+(x)$,  line 
(c) and ($\bullet$): $p_{s,1}^-(x)$.}
\label{Fig10}
\end{figure}

In order to derive the proper boundary conditions it is
therefore convenient to consider the restriction of eq. (\ref{eq6_2})
to each of the two invariant intervals $I_h$, $h=1,2$.
Consider eq. (\ref{eq6_2}) restricted to $x \in I_1=[0,x^*]$.
At $x=0$, the usual reflective condition applies. 
At the other endpoint $x=x^*$ of the invariant interval,
 zero net-flux 
condition should be enforced, namely,
\begin{equation}
\left [ v(x^*)+c \right ] \, p^+(x^*,t) + \left [v(x^*)-c \right ] \, p^-(x^*,t)
= 0 \; .
\label{eq6_9}
\end{equation}
Since $ v(x^*)+c =0$, eq. (\ref{eq6_9}) implies
\begin{equation}
p^-(x^*,t) = 0 \; ,
\label{eq6_10}
\end{equation}
which indicates that the regressive wave should vanish at $x^*$.
Figure \ref{Fig10} depicts the comparison of the
PPM  (\ref{eq6_1}) restricted to $I_1$ (solid lines)
equipped with the boundary condition (\ref{eq6_10})
and the results of stochastic simulation for $p_{s,1}(x)$
and its partial components $p_{s,1}^+(x)$, $p_{s,1}^-(x)$ at
$\nu=1.1$. An ensemble of $N=10^6$ particles has been used
in the stochastic simulations. As expected from
eq. (\ref{eq6_10}), stochastic simulation data confirm the vanishing
value of the regressive wave at $x^*$.

Consider $x \in I_s=[x^{**},1]$. Similarly to eq. (\ref{eq6_10})
one derives
\begin{equation}
p^+(x^{**},t)=0 \; ,
\label{eq6_11}
\end{equation}
since $v^-(x^{**})=v(x^{**})-c=0$.

As  stated above, the boundary condition at $x=1$ does not derive from
a balance between partial waves (as both are progressive waves 
in the neighbourhood of $x=1$), but is
a consequence of the assumption of an infinite potential
barrier preventing particles to escape from the boundary in stochastic 
simulations.

This observation implies that the partial waves possess
an absolutely continuous component $\widetilde{p}^{\pm}(x,t)$
and an impulsive part localized at $x=1$,
\begin{equation}
p^{\pm}(x,t)= \widetilde{p}^{\pm}(x,t)+ \pi^{\pm}(t) \, \delta(x-1) \; ,
\label{eq6_12}
\end{equation}
where the values of $\pi^{\pm}(t)$ stem from probability conservation.
Asymptotically, a steady impulsive distribution is
obtained, $p_{s,2}(x)=\delta(x-1)$, as physically
expected since $v^\pm(x) \geq 0$ uniformly in $I_2$, as can be observed in figure \ref{Fig9} in panels $(a)$ and $(c)$.

\section{Concluding remarks}
\label{sec7}

In this article we have analyzed in detail the structure
and the admissibility of different forms of boundary
conditions for one-dimensional
hyperbolic transport problems on the interval,  both in the absence and
in the presence of a deterministic biasing field.

We stress again that the concept of a
  {\em hyperbolic transport model}  should be regarded as  a
system  of  balance equations,
obtained from the underlying micro-dynamics
 expressed in the form of a stochastic differential equation
driven by dichotomous noise and its generalizations.
In this framework, hyperbolic transport  theory is the wave-like counterpart
of  the classical transport approach in which the stochastic micro-dynamics is
controlled by a Langevin equation driven by
Wiener processes (Brownian motion).

In this concluding Section we would  like to pinpoint some
general observations emerging from the present work that
can be used to develop a self-consistent theory of hyperbolic
transport, not limited to one-dimensional problems, and
that be will thoroughly elaborated elsewhere:

\begin{enumerate}
\item The primitive  observables in hyperbolic
transport theory
are the partial probabilities (in an one-dimensional model with
a single velocity $c$, these are just the partial probability
waves $p^+(x,t)$ and $p^-(x,t)$). The existence of these
observables derives intrinsically from the dichotomous nature of
the underlying stochastic fluctuations at microscale.
\item The overall concentration $p(x,t)$ as well as its flux $J(x,t)$ 
should be
regarded in the theory as {\em derived} quantities, although they
represent the physical observables commonly measured in transport experiments.
This result is somehow similar to the setting of non-relativistic
 quantum mechanics
where the wave function  is the primitive quantity
of the theory, while the probabilistic description of  a quantum system
is associated to the square of its modulus (Born condition).
\item  Boundary conditions in the hyperbolic theory are written
as relations connecting forward and backward probability waves, corresponding
to transmission/reflection conditions for the partial probability
waves. The set of admissible boundary conditions
is dictated by the positivity of partial probability functions,
and can be derived alternatively by invoking the dissipative
nature of the dynamics (Section \ref{sec5}).
\item There is a close connection between the functional
relation describing 
dissipation in hyperbolic ``diffusion'' eq. (\ref{eq5_3})
and the approach followed in {\em Extended Irreversible Thermodynamics}
\cite{eit1,eit2,eit3},
where the entropy function depends also on the flux and not
only on the concentration.
The observation that the basic primitive quantities
in hyperbolic transport theory are the partial waves suggests
that the thermodynamic formalism developed in
the theory of {\em Extended Irreversible Thermodynamics}
could be reformulated and/or generalized 
in order to include the primitive variables
of hyperbolic transport theory (namely the partial probabilities)
as the system variables with respect to which the state functions
(such as entropy) are expressed. This effort would permit
to resolve the problem of the absence of any probabilistic
meaning of the Cattaneo equation (which is the prototype
of generalized transport equation with memory in extended
thermodynamics) in space dimensions greater than one.
\end{enumerate}



%% References


\bibliographystyle{model1-num-names}
\bibliography{<your-bib-database>}



\end{document}